\documentclass[sigconf]{acmart}

\usepackage{blindtext}
\usepackage{xcolor}
\usepackage{graphicx}
\usepackage{subcaption}
\usepackage{booktabs}
\usepackage{tcolorbox}
\usepackage{blindtext}
\usepackage{graphicx}
\usepackage{tikz}
\usepackage{listings}
\usepackage{pgfplots}
\usepackage{pgfplotstable}
\usepackage{float}
\usepackage{tabularx}
\usetikzlibrary{positioning, arrows.meta, shapes.geometric, fit}
\usepackage{soul}

\newcommand{\code}[1]{%
  {\setlength{\fboxsep}{1.5pt}
  \setlength{\fboxrule}{0pt}
  \raisebox{0pt}[.5\height][0.1\depth]{
    \fbox{\colorbox{gray!20}{\texttt{#1}}}}%
  }%
}

\newcommand{\myparagraph}[1]{\noindent\textbf{#1:}}

\AtBeginDocument{%
  }

\setcopyright{rightsretained}
\copyrightyear{2025}
\acmYear{2025}
\acmConference[ITiCSE 2025 (to appear in)]{Proceedings of the 2024 Innovation and Technology in Computer Science Education}{June 30--July 2, 2025}{Nijmegen, The Netherlands}
\acmBooktitle{Proceedings of the 2025 Innovation and Technology in Computer Science Education (to appear in ITiCSE 2025), June 30--July2, 2025, Nijmegen, The Netherlands}
\acmDOI{XXXXXXX.XXXXXXX}

\begin{document}

\title[ReDefining Code Comprehension]{ReDefining Code Comprehension: Function Naming as a Mechanism for Evaluating Code Comprehension}

\author{David H. Smith IV}
\email{dhsmith2@illinois.edu}
\orcid{0000-0002-6572-4347}
\affiliation{%
  \institution{University of Illinois}
  \city{Urbana}
  \country{USA}
}

\author{Max Fowler}
\email{mfowler5@illinois.edu}
\orcid{0000-0002-4730-447X}
\affiliation{%
  \institution{University of Illinois}
  \city{Urbana}
  \country{USA}
}

\author{Paul Denny}
\email{p.denny@auckland.ac.nz}
\orcid{0000-0002-5150-9806}
\affiliation{%
  \institution{University of Auckland}
  \country{New Zealand}
}

\author{Craig Zilles}
\email{zilles@illinois.edu}
\orcid{}
\affiliation{%
  \institution{University of Illinois}
  \city{Urbana}
  \country{USA}
}

\renewcommand{\shortauthors}{}

\begin{abstract}

"Explain in Plain English" (EiPE) questions are widely used to assess code
comprehension skills but are challenging to grade automatically. Recent
approaches like Code Generation Based Grading (CGBG) leverage large language
models (LLMs) to generate code from student explanations and validate its
equivalence to the original code using unit tests. However, this approach does
not differentiate between high-level, purpose-focused responses and low-level,
implementation-focused ones, limiting its effectiveness in assessing
comprehension level. We propose a modified approach where students generate
function names, emphasizing the function's purpose over implementation details.
We evaluate this method in an introductory programming course and analyze it
using Item Response Theory (IRT) to understand its effectiveness as exam items
and its alignment with traditional EiPE grading standards. We also publish this
work as an open source Python package for autograding EiPE questions, providing
a scalable solution for adoption.

\end{abstract}



\maketitle

\section{Introduction}

As we continue to transition from the age of traditional programming to the new
age of Human-GenAI collaborative coding, it is essential that we prepare
students with the skills necessary to succeed in such collaborations. In this
transition from traditional CS1 to CS1-LLM~\cite{vadaparty2024cs1}, instructors
have begun to discuss a possible shift towards emphasizing the skills of
describing computational tasks through natural language~\cite{reeves2024prompts,
denny2024prompt} and comprehending code produced by an LLM to verify its
correctness~\cite{prather2023robots, prather2024beyond}. One existing approach
for assessing these skills are ``Explain in Plain English'' (EiPE) questions, in
which students describe the purpose of a given code
snippet~\cite{fowler2021should}. However, their adoption has been limited by the
time-intensive grading process of marking short answer responses.

Though prior autograding approaches developed for EiPE questions have seen some
success, they each suffer from distinct limitations. A logistic classifier
developed by \citet{fowler2021autograding} matches the performance of a trained human grader but requires a large corpus of human labeled data for the creation of each question.
To overcome these drawbacks,
\citet{smith2024code} proposed Code Generation Based Grading (CGBG), an
autograding approach that uses a large language model (LLM) to generate code
from a student's description and verifies its functional equivalence to the code the student was describing with unit
tests. 
However, this approach struggles to distinguish between high-level, purpose-focused, and low-level,
implementation-focused code descriptions.

To address the limitations of prior approaches, we propose a modified approach to
CGBG EiPE questions where, rather than having students provide an unbounded
description of code, we ask that they provide a syntactically valid Python
function name. We hypothesize that the more bounded nature of the entry will nudge students towards high-level descriptions. 
To evaluate this approach we explore the following research questions:
\begin{enumerate}
  \item[\textbf{RQ1:}] What are the psychometric properties of function
    redefinition EiPE questions when used in an exam setting?
  \item[\textbf{RQ2:}] What is the alignment between the autograding mechanism
    for function redefinition EiPE questions and the grading standards of
    standard EiPE questions?
\end{enumerate}
Additionally, to allow for the easy adoption of this grading approach, we
release it as a new feature in the \texttt{eiplgrader}\footnote{https://github.com/CoffeePoweredComputers/eiplgrader} package,
an open source Python package for autograding EiPE questions.

\section{Background}

To contextualize this work we first provide an overview of Explain in Plain
English (EiPE) questions in Section~\ref{sec:traditional-eipe}, discussing the
their importance and approaches to grading them. In
Section~\ref{sec:eipe-autograding} we discuss autograding approaches relevant to
EiPE questions along with their advantages and limitations. 

\subsection{Traditional EiPE Questions}\label{sec:traditional-eipe}

\begin{figure}[H]
    \centering
    \begin{tikzpicture}
        \draw[rounded corners, fill=blue!5] (0,0) rectangle (8,-5);

        \node[anchor=west, text=black!70, font=\large\bfseries] at (0.2, -0.5) {Describe the following code:};

        \node[anchor=north west, text=black, rounded corners] at
        (0.5, -0.75) {
            \begin{minipage}{6.75cm}
              
              \begin{lstlisting}[]
def foo(a, b):
    y = 0
    for e in a
        if e == b
            y += 1
    return y
                \end{lstlisting}
            \end{minipage}
        };

        \draw[rounded corners, fill=white, draw=black] (0.5, -3.5) rectangle
        (7.5, -4.75);
        \node[anchor=north west] at (0.7, -3.6) {
            \textbf{Your Answer:}
        };
        \node[anchor=north west, text=gray] at (0.7, -4) {
            \textit{Enter your response here...}
        };
    \end{tikzpicture}
    \caption{Sample interface for an Explain in Plain English (EiPE) question.}\label{fig:eipe_question}
\end{figure}

Explain in Plain English (EiPE) questions (Figure~\ref{fig:eipe_question}) are a
common approach for assessing and developing code comprehension skills. In these
questions, students are shown a segment of code and asked to provide a
high-level, natural language description of what that code is doing. 

With respect to what differentiates ideal from suboptimal EiPE responses, the
``Structure of Observed Learning Outcome'' (SOLO)
taxonomy~\cite{biggs2014evaluating} has often been used to categorize responses
based on their demonstrated level of comprehension. An adapted version of this
taxonomy, introduced by \citet{lister2006not}, defined the following levels of
comprehension for EiPE responses: 
\begin{itemize}
  \item \textbf{Relational:} The student's response demonstrates an
    understanding of the relationships between the code's components by
    describing the code's purpose rather than its implementation. 
  \item \textbf{Multistructural:} The student's response demonstrates an
    understanding of the code's components but not the relationships between
    them. This often manifests as line-by-line descriptions of the code.
  \item \textbf{Unistructural:} The student's response demonstrates an
    understanding of some components of the code but is either missing 
    components or contains errors.
  \item \textbf{Prestructural:} The student's response does not contain
    any relevant or correct descriptions of the code.
\end{itemize}
\citet{clear2008reliably} further refined this taxonomy by introducing the
categories of \textit{Relational Error} and \textit{Multistructural Error} to
account for responses that demonstrate a high-level understanding of the code's
purpose or provide a full description of the code, respectively, but contain
errors.

Prior work on the topic of EiPE---and code comprehension more generally---has
indicated a correlation between students who demonstrate high-level code
comprehension skills and are able to ``see the forest for the
trees''~\cite{lister2006not} and students who are able to succeed on code
writing tasks~\cite{fowler2022reevaluating, lopez2008relationships,
lister2009further}. However, eliciting high-level responses may not always be straightforward.  For example, Sheard et al. hypothesized that some multistructural responses may stem from a misunderstanding of the instruction to ``explain in plain English'', or to ``explain the purpose'' of code~\cite{sheard2008going}.  To address this, they suggested that asking students to nominate a function name might be a promising alternative, as a way of guiding them towards relational thinking. However, to our knowledge, this approach has not been studied in a research setting. Additionally, prior studies have highlighted that expressing ideas clearly in English can be challenging for some students, particularly in linguistically diverse contexts, where language issues may affect their ability to articulate high-level responses~\cite{kumar2021refute}.

In proposing their theory of instruction for introductory programming skills, Xie et al.  \cite{xie2019theory} suggest that code
comprehension is a critical skill, and one that should be taught immediately
prior to code writing to improve success in those tasks. In the context of GenAI assisted programming, many have suggested that
the ability to articulate computational tasks through natural language and
comprehend code will play more critical roles in computing education given
students given the alignment between these skills and the skills of prompting
and verifying code produced by an LLM~\cite{reeves2024prompts, denny2024prompt,
prather2023robots, prather2024beyond}.

\subsection{Auto-grading EiPE Questions}\label{sec:eipe-autograding}

Despite the clear importance and utility of EiPE questions, their adoption has
been limited by the time-consuming nature of grading free-response
questions~\cite{fowler2021should}. To address this challenge, several
auto-grading approaches have been developed with some success. 

\citet{fowler2021autograding} developed an logistic classifier trained on a
large corpus of human-labeled responses for each question. This approach has
been shown to provide grades that reliably match those of a trained teaching
assistant. However, it is limited by two factors. First, there is a significant
question authoring overhead as a large corpus of human-labeled responses must be
created \textit{for each question}. Second, though the core goal of providing
automated feedback was achieved, the feedback was limited to marking answers as correct or incorrect while providing some sample correct answers.

To address these limitations, \citet{smith2024code} proposed an autograding
approach---Code Generation Based Grading (CGBG)---that leverages a large
language models to generate code from a student's description and uses a suite
of unit tests to verify the code generated is functionally equivalent to the
code the student was describing. Several evaluations of this approach have
demonstrated its effectiveness and as well as its ability to provide students
with feedback through the results of unit-tests and the generated
code~\cite{denny2024explaining, smith2024prompting, kerslake2024integrating}.
Additionally, given modern LLMs are trained on a wide variety of written
languages, this approach has been shown to allow for the autograding of EiPE
questions in a variety of languages~\cite{smith2024explain, smith2024prompting}.
However, a limitation of this approach is that it is unable to distinguish
between low-level responses and high-level ones, limiting its effectiveness at
evaluating comprehension level.

\begin{table*}
  \centering
  \caption{Question IDs and their corresponding code which students were asked to describe}\label{tab:qids_code}
  \vspace{-0.25cm}

\begin{tabular}{@{}lcccc@{}}
  \toprule
  \textbf{Question} & Count Odd Nums in List & Get Numbers Below Threshold & Absolute Values of List & Count Strings of Given Length \\
  \textbf{Code} &
  \begin{lstlisting}[
    language=Python,
    basicstyle=\small,
    keywordstyle=\color{purple},
  ]
def foo(x: List[int]):
    k = 0
    for e in x:
        if e % 2 != 0:
            k += 1
    return k
  \end{lstlisting} &
  \begin{lstlisting}[
    language=Python,
    basicstyle=\small,
    keywordstyle=\color{purple},
  ]
def foo(x: List[int], t: int):
    result = []
    for e in x:
        if e < t:
            result.append(e)
    return result
  \end{lstlisting} &
  \begin{lstlisting}[
    language=Python,
    basicstyle=\small,
    keywordstyle=\color{purple},
  ]
def foo(x: List[int]):
    for i in range(len(x)):
        if x[i] < 0:
            x[i] = -x[i]
  \end{lstlisting} &
  \begin{lstlisting}[
    language=Python,
    basicstyle=\small,
    keywordstyle=\color{purple},
  ]
def foo(x: List[str], n: int):
    k = 0
    for s in x:
        if len(s) == n:
            k += 1
    return k
  \end{lstlisting} \\ 
  \bottomrule
\end{tabular}

\end{table*}

\section{Methods}

\begin{figure}
  \centering
  \includegraphics[width=\columnwidth]{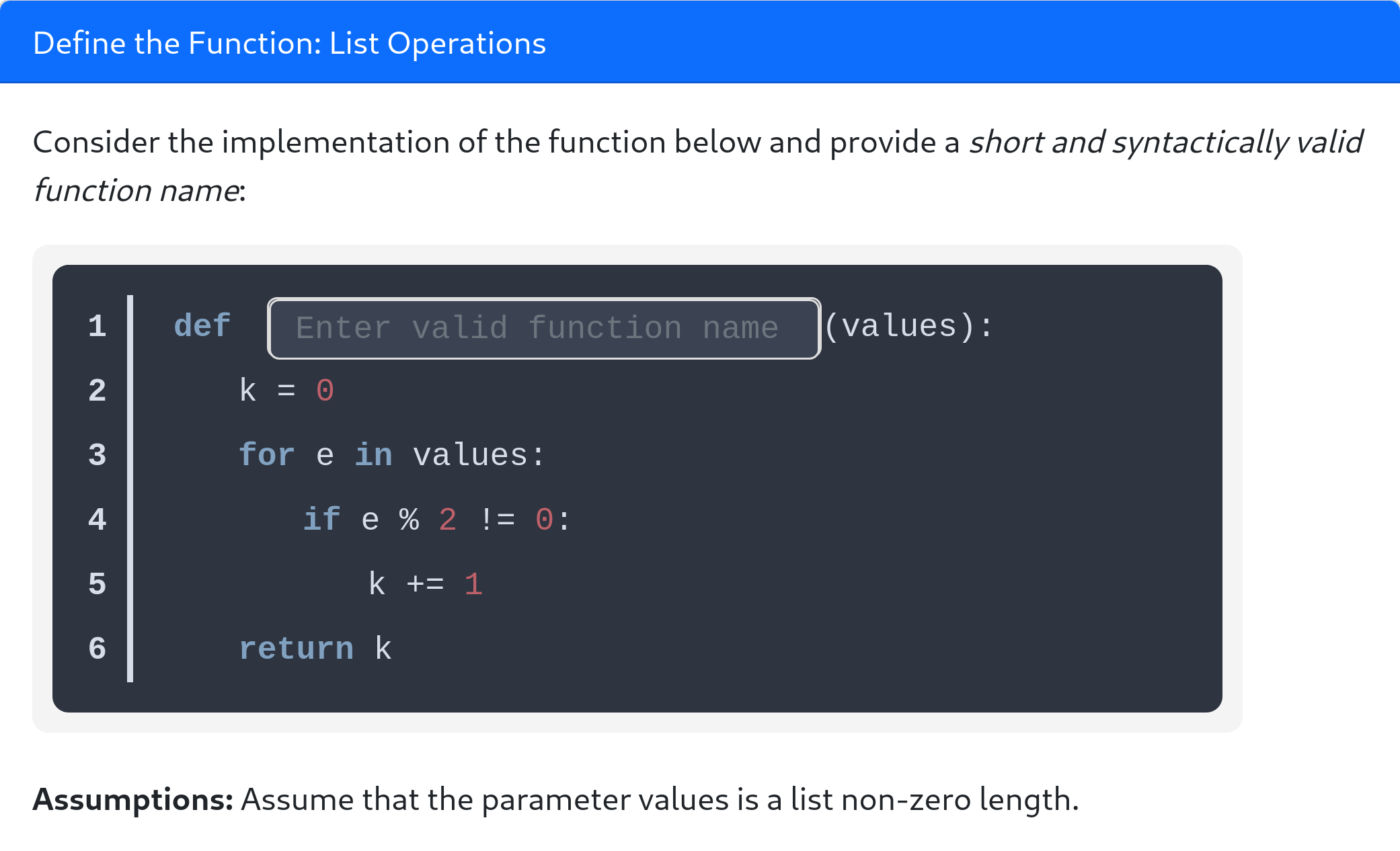}
  \vspace{-0.5cm}
  \caption{The interface of the function redefinition EiPE task.}\label{fig:func_interface}
  \vspace{-0.25cm}
\end{figure}

\begin{figure*}
    \centering
    \includegraphics[width=\textwidth]{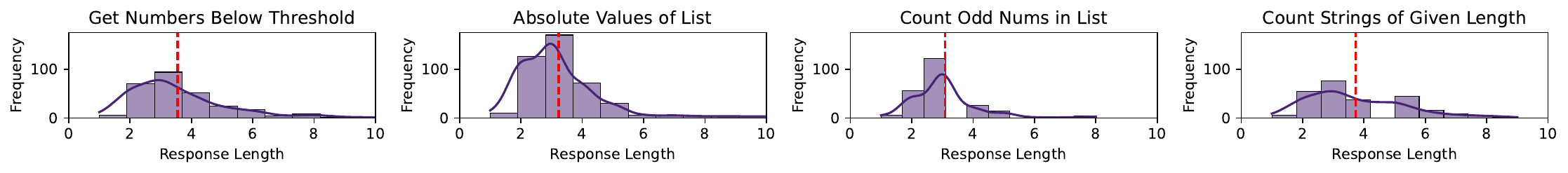}
    \caption{Lengths of all responses (N=647) submitted for each question}\label{fig:length}
\end{figure*}

\begin{figure*}
    \centering
    \includegraphics[width=\textwidth]{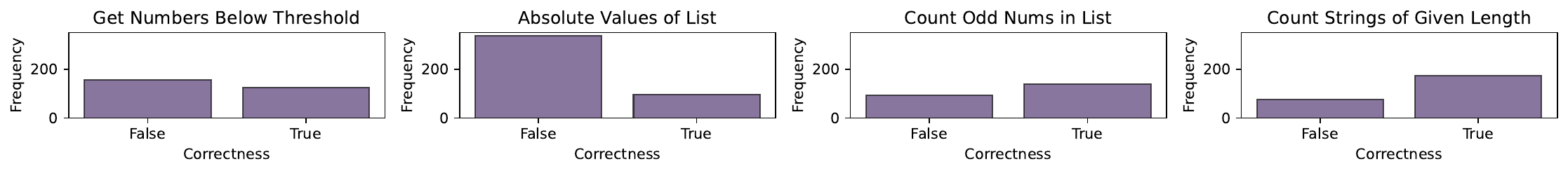}
    \caption{Correctness of all responses (N=647) for each question.}\label{fig:correct}
\end{figure*}

To collect student responses for the evaluation of Function Redefinition EiPE
questions (Figure~\ref{fig:func_interface}), we constructed four questions.
Students (N=366) from a large, introductory programming course in the
United States were each randomly assigned one of these questions on
a final exam (Table~\ref{tab:qids_code}). The exams took place in a proctored
computer lab~\cite{zilles2019every} and were administered on the PrairieLearn
platform~\cite{west2015prairielearn}.

Responses to these questions had two restrictions: the student's response must
be a syntactically valid Python function name and must be no longer than 10
words. In the event a student's submission did not meet these standards they
were informed via an error message and given the opportunity to resubmit
without penalty. Students were given three attempts to get the questions
correct, each time receiving feedback on the correctness of their response
through the standard CGBG approach. In total, we collected 647 responses of
which 476 were unique.

To investigate the alignment between the Function Redefinition EiPE questions
and the objectives of EiPE questions, we perform an analysis
of all student responses collected during the exam. In doing so, we re-grade
each response using two approaches:
\begin{itemize}
  \item \textbf{One Attempt Grading:} Here, the students response is used to generate a single Python function. This function is then executed on a set of test cases to determine if the function is correct. 
  \item \textbf{Robustness Grading:} In this approach, the student's response is used to generate a set of 5 Python functions. These functions are then executed on a set of test cases to determine if each function is correct. The response is considered correct if \textit{all} of the generated functions passes \textit{all} test cases.
\end{itemize}
Responses were generated using OpenAI's GPT-4o model with a temperature of 0 to 
ensure deterministic responses. 

We compare the outcomes with respect to the psychometric properties of the questions
as they were graded under each of these approaches (Section~\ref{subsec:irt}) and
the analysis of responses using the SOLO taxonomy
(Section~\ref{subsec:solo-analysis}) which are discussed in the following
subsections.

\section{Results}

We first look at descriptive statistics for each question---namely response length and correctness--to gain a high-level understanding of student responses to each of the questions.

As expected, due to the 10-word limit imposed by the function redefinition
exercises no responses exceeded this limit. The vast majority of the responses
were far shorter, with none of the questions exceeding four words as their
average length (Figure~\ref{fig:length}).

As for performance, there was significant variance in the apparent difficulty
of each question (Figure~\ref{fig:correct}). \textit{Count Strings of Given
Length} and \textit{Count Odd Nums in List} students submitted more
correct rather than incorrect responses. In contrast, \textit{Get Numbers Below
Threshold} was more difficulty with the proportion of correct responses
only slightly exceeding the proportion of incorrect responses. Finally,
\textit{Absolute Values of List} was by far the most difficult with the
proportion of incorrect responses far exceeding those that were graded as
correct.
\subsection{RQ1) Item Response Theory}\label{subsec:irt}

\begin{figure}
  \centering
  \begin{subfigure}[b]{0.49\textwidth}
    \centering
    \includegraphics[width=\textwidth]{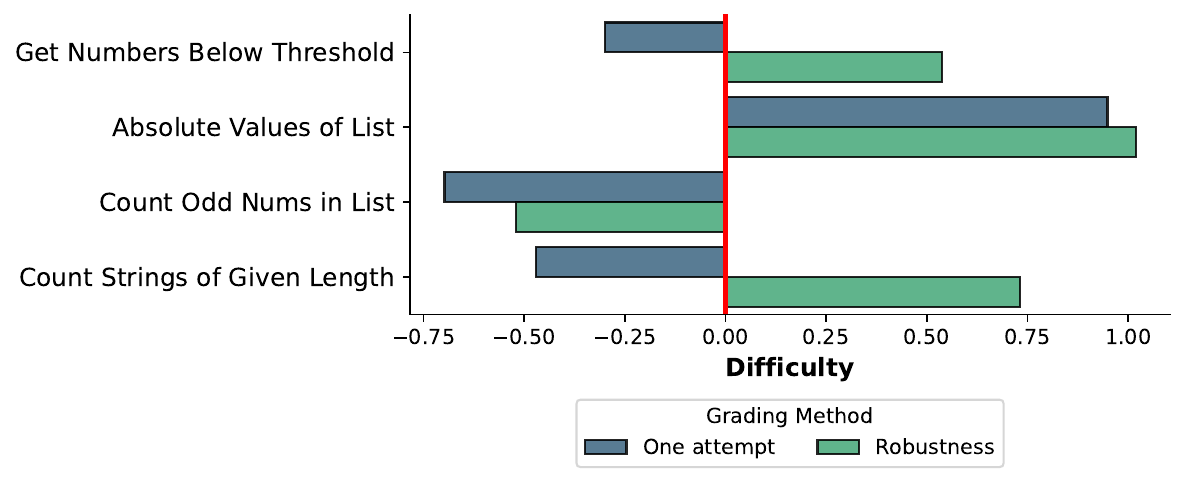}
    \caption{Difficulty ($b$) Coefficients}
    \label{fig:diff}
  \end{subfigure}
  \begin{subfigure}[b]{0.49\textwidth}
    \centering
    \includegraphics[width=\textwidth]{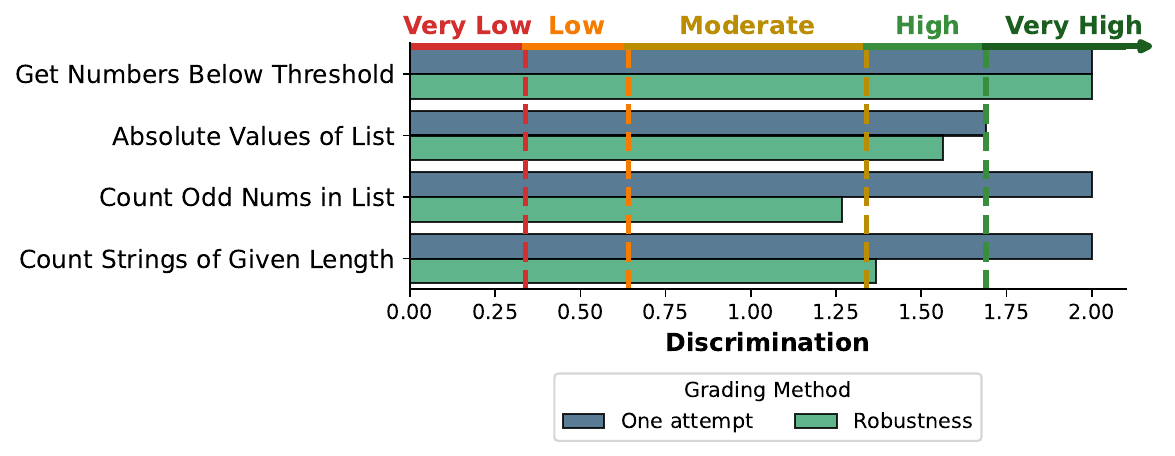}
    \caption{Discrimination ($a$) Coefficients}
    \label{fig:discrim}
  \end{subfigure}
  \caption{Item Statistics from the Results of Fitting 2PL IRT}
\end{figure}

For our analysis we employ a two parameter logistic (2PL) item response (IRT)
model~\cite{birnbaum1968some}.
\[ P(\theta) = \frac{1}{1 + e^{a_i(\theta-b_i)}} \]
We use this model to predict the probability of a student of a given ability
level $\theta$ responding correctly to an item. The item statistics are as
follows:
\begin{enumerate}
  \item[$a_i$] The discrimination of item $i$. Discrimination is the slope of the
    logistic curve at the intercept with a 50\% probability of a correct
    response and characterizes the ability of the item to discriminate between
    students above and below that ability level.
  \item[$b_i$] The difficulty of item $i$. Difficulty is characterized as the
    ability level ($\theta$) at which a student has a 50\% chance of responding
    to a question correctly.
\end{enumerate}
A shortcoming of traditional 2PL is that it assumes all questions were
dictomously graded---an assumption that does not hold with the data collected
in this study given that students were allowed multiple attempts and awarded
partial credit through a variety of mechanisms (e.g., partially passing unit
tests). To avoid the potential loss of information that we may incur through an
alternative method of dicotomizing---such as rounding or only considering the
first attempt---we use an adapted version of the cross entropy optimization
approach introduced by \citet{chen2024plagiarism}. In keeping with their
approach we also adopt the standard bounds of each of the model's coefficients:
$\theta \in [-3, 3]$, $a \in [0, 2]$, $b \in [-3, 3]$. We discuss the results of
this analysis in the following subsection. Additionally, to aid in the interpretation of 
the discrimination coefficients we use the cutoffs presented by
\citet{baker2001basics} of: Very Low ($\leq0.34$), Low ($0.35-0.64$), Moderate
($0.65-1.34$), High ($1.35-1.69$), and Very High ($>1.7$).

\subsubsection{Results}

Overall, the results of the IRT analysis indicate that the Function Redefinition
EiPE questions function well as exam items, exhibiting high discrimination
(Figure~\ref{fig:discrim}) with all questions exceeding the \textit{moderate}
threshold and many exceeding \textit{high} and \textit{very high}. In terms of
differences in these outcomes between the two grading approaches, we observe
that in three cases questions graded under the \textit{\textbf{One Attempt}} approach
exhibited a higher discrimination coefficient than those graded under the
\textit{\textbf{Robustness}} approach. In two cases, this difference was substantial,
lowering the coefficient from well exceeding the \textit{very high} threshold to
the upper end of \textit{moderate} and the lower end of \textit{high}.

The primary differences between the \textit{\textbf{One Attempt}} and
\textit{\textbf{Robustness}} grading approaches appears with respect to the
difficulty (Figure~\ref{fig:diff}). In the \textit{\textbf{One Attempt}}
approach, the difficulty of of the questions is relatively low, with only one
question---\textit{Absolute Values of List}---achieving a positive difficulty
coefficient. Conversly, under the \textit{\textbf{Robustness}} approach, all
questions except one---\textit{Count Odd Nums in List}---exhibit a positive
difficulty coefficient. 

\vspace{2mm}
\noindent
\setlength{\fboxsep}{10pt} 
\setlength{\fboxrule}{0.5pt} 
\fcolorbox{black}{gray!10}{
\parbox{\dimexpr\linewidth-2\fboxsep-2\fboxrule\relax}{
\textbf{Key Takeaways:} These findings indicate that both approaches produce
items that effectively discriminate between students of different ability
levels. The primary consideration for instructors is the trade-off between
difficulty and discrimination. The \textit{\textbf{One Attempt}} approach produces items
that are easier for students, while the \textit{\textbf{Robustness}} approach produces
items that are more difficult with a small reduction in discrimination. 
}}

\subsection{RQ2) Alignment with Traditional EiPE}\label{subsec:solo-analysis}

To analyze the quality of the responses to the Function Redefinition EiPE
questions---and by extension their alignment with the objectives of EiPE
questions---we conducted a qualitative analysis of the responses using the
modified SOLO taxonomy presented by \citet{clear2008reliably}. Two researchers
independently coded the responses into the categories of \textit{Relational},
\textit{Relational Error}, \textit{Multistructural}, and
\textit{Multistructural Error}, or \textit{Other Error}. Agreement was
calculated using Cohen's $\kappa$, which was found to be acceptable at a value
of 0.75, and any disagreements were resolved through discussion. 

\subsubsection{Results}

\begin{figure*}
  \centering
  \includegraphics[width=\textwidth]{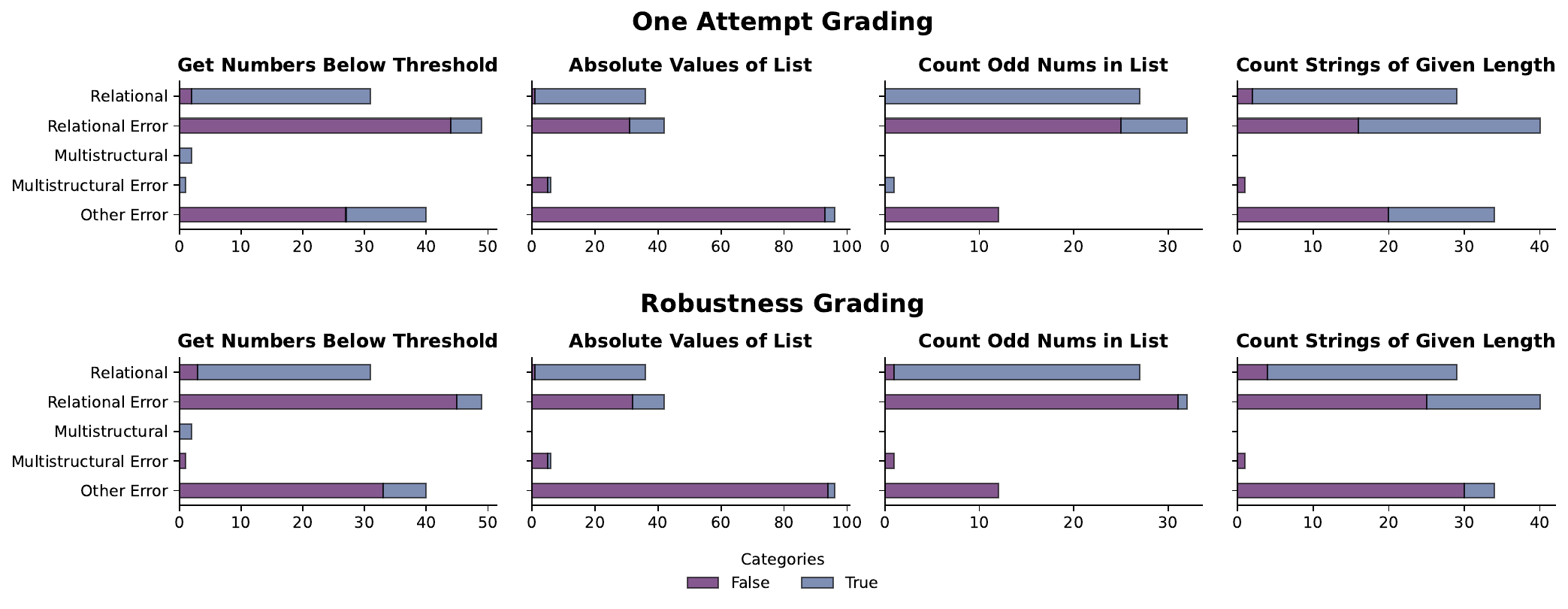}
  \caption{The percentage of responses belonging to each SOLO taxonomy category graded as correct for each grading approach.}
  \label{fig:redef_eipe}
\end{figure*}

Overall, the results of the comparison between the Function Redefinition EiPE
autograding approach and the SOLO taxonomy indicate that the function redefinition
often aligns with the objectives of EiPE questions. Across all questions---and in
both grading approaches---the vast majority of responses were categorized as
\textit{Relational} were graded as correct. Additionally, given the near complete absence of \textit{Multistructural} responses it would appear that the question format was successful in dissuading students from providing such responses.

Looking at the comparison of the two grading approaches, all cases, the use of
the \textit{\textbf{Robustness}} grading approach resulted in a higher proportion of
\textit{Relational Error} and \textit{Other Error} responses being graded as
incorrect, thus reducing the prevalence of false positives, while having a very
minimal impact in terms of increasing the prevalence of false negatives (i.e.,
\textit{Relational} responses graded as incorrect). 

To characterize the alignment and misalignments that exist between
this activity and its grading approaches, we wish to highlight two categories:
1) relational responses graded as correct and 2) relational errors and other errors graded as correct. In doing so, we aim 
to characterize the nature of what we regard as true positives and false positives due to their misalignment with the SOLO taxonomy. \\

\myparagraph{Relational Responses Graded as Correct}
\textit{Relational} responses graded as correct included
\code{get\_values\_under\_threshold} and \code{count\_odd\_nums}. Such responses
are relational and precise, as they describe the action being performed by the
function (e.g., getting, counting), the property of the object being acted
upon (i.e., values, nums), and, in the case of the former, the condition under
which the action is performed (i.e., below a threshold). Such responses were
consistently graded as correct under both the \textit{\textbf{One Attempt}} and
\textit{\textbf{Robustness}} grading approaches.\\

\myparagraph{Relational Errors and Other Errors Graded as Correct}
In contrast to \textit{Relational} responses, \textit{Relational Error}
responses which were graded as correct often missed details. For example, with
regard to the \textit{Count Strings of Given Length} question, the response
\code{count\_strings} is correct in that it describes the action being performed
but fails to provide the necessary context of what is being counted. Similarly a
response from the \textit{Other Error} category which was graded as correct,
\code{count\_same\_length} is correct in that it describes the action being
performed, but it fails to provide the necessary context of what is being
counted and what is meant by \code{same}. These absences make it difficult to
determine the purpose of the function without additional context and, as such,
do not offer a clear indication that the student providing the response has a
complete understanding of the code that they are describing. The use of the
\textit{\textbf{Robustness}} grading approach did successfully filter out the
first example but not the second. These results suggest that the combination of
the student-provided function name along with the question author provided
parameters and assumptions was sufficient for the LLM to accurately ``guess''
the correct code.

Such good guesswork on the part of the LLM was not limited to this question and
appeared to consistently be the reason for questions in \textit{Other Error} and
\textit{Relational Error} categories being graded as correct. Other examples
include: \code{less\_than\_t} for \textit{Get Numbers Below Threshold} and
\code{find\_distance\_to\_zero} for \textit{Absolute Values of List}.

\vspace{2mm}
\noindent
\setlength{\fboxsep}{10pt} 
\setlength{\fboxrule}{0.5pt} 
\fcolorbox{black}{gray!10}{
\parbox{\dimexpr\linewidth-2\fboxsep-2\fboxrule\relax}{
\textbf{Key Takeaway:} Overall, the Function Redefinition EiPE questions exhibit
a high degree of alignment with the objectives of EiPE questions and appears to
have deterred the use of multistructural responses. Of the two grading
approaches evaluated, the \textit{\textbf{Robustness}} approach proved an essential---though
admittedly not foolproof---mechanism for mitigating the
risk of false positives while having a minimal impact on true positives. 
}}

\section{Discussion}

Function naming exercises present an effective and exciting way to assess students' ability to comprehend code. Given the constrained response, compared to traditional EiPE, the question type almost totally avoids multistructural responses, keeping students' answer efforts more relational. In terms of summative assessment, the exercises have at least moderate, and generally high, discrimination, providing value as test items. While there is a trade-off in difficulty behind whether one chooses to use one-attempt or robustness grading, using the latter reduces the risk that GPT-4o, capable as it is, constructs successful functions from answers with small errors. This reduces the number relational errors marked correct by the grader, with only \textit{Count Strings of Given Length} showing less success. We attribute this difference in performance to the structure of the question. Considering an example relational error \code{count\_strings}, our grading paradigm and the student's answer provides the action (counting), data types (strings), function definition (count\_strings), and question assumptions (the question takes a list and an integer). This may be a more trivial problem for an LLM to fill in the blanks to generate successful functions more often, if only because the integer parameter implies some actions with that integer, for which a comparison with string lengths is one of the few reasonable assumptions the model could make. 

In light of function naming exercises constraining students more towards relational responses, we are particularly excited by the potential for these exercises to serve as scaffolding for building students' competence for traditional EiPE exercises and, broadly, the skills of comprehending and explaining pieces of code. It may be that having to describe a piece of code in a few key words is a useful precursor to being able to reason about and describe the code at a high-level of abstraction. Summatively, this could provide a more gentle curve of students' code comprehension ability: rather than not at all or mastering EiPE, it at least provides an intermediate skill of appropriate function naming. Formatively, it may be that function naming is a more accessible task earlier in the semester of a programming course and, as students develop their programming and code comprehension skills, function naming tasks can lead into more traditional, high-level EiPE tasks. One avenue for future work could explore is using the same functions for function naming exercises and traditional EiPE exercises to directly compare difficulty between the two and further investigate this potential relationship between the skills.

The robustness grading method for these questions has additional potential benefits beyond reducing false positives in the grading process. Robustness grading may itself prove to be a valuable partial credit mechanism, which is an open question that plagues the use of EiPE exercises~\citep{fowler2021should}. Students can receive points not just for generating a correct function, but for how many correct functions their proposed function name generates. This can still be automated with the same test cases, but allows for more granular scoring and feedback to students. Function names that work five out of five times are likely higher quality than function names that happen to work once out of five times, which itself is more feedback than all-or-nothing correctness grading. This may also have the benefit of making students happier, as some points are better than no points.

Finally, we wish to briefly comment on the large number of relational errors we noted in students' responses, especially in light of the false positive count for \textit{Count Strings of Given Length}. In applying the SOLO taxonomy, our standard was arguably maximally harsh in that we leveraged our existing knowledge of applying the taxonomy to \textit{full EiPE responses} and only considered the function names, not the other assumptions and information available. While EiPE responses are ideally short, the size of a function is shorter still. The function name \code{count\_strings} may imply something along the lines of ``a function that counts all the strings in a mixed-type list,'' in a vacuum. However, in the context of knowing the function takes a list of strings and an integer, it may be reasonable to take the name \textit{and} the arguments as sufficient to mark the answer as relational \textit{without} an error. Other researchers may have ended up with a lower count of relational errors with a less harsh yet still reasonable standard. There may also be value in adapting the SOLO taxonomy for use with function name sized responses, especially in light of multistructural responses barely existing.

\section{Limitations}
There are some key limitations that we wish to note. First, while we consider \textit{Count Strings of Given Length} to be an outlier that performs more poorly than our other questions, it is also possible that the other three questions may instead be outliers that perform unusually well for this task. We cannot be certain our findings generalize to all code used for function naming exercises. Future work that investigates more code snippets will be useful for refining our understanding of the exercises' difficulty and discrimination, i.e. their value as test items.

There are two course specific limitations we wish to raise. First, students in the class saw these exercises relatively late into the semester: for the first time on homework during week 12 of a 16 week course. This means that students did not have the whole semester to learn how to work with these questions and students' performance will likely differ when given longer periods of time to work with these styles of questions. On the other hand, students in this course are exposed to traditional EiPE exercises regularly from the first week of the class onward, with exercises focused on small functions being commonplace from the fourth week onward. Given this, the students in this course may have been more suited to function naming exercises than typical CS0/CS1 students. In turn, studies with different student populations may find higher difficulty measures than we found here if their student population is less familiar with code comprehension being assessed.

\section{Conclusion}

We propose and evaluate function naming exercises as an approach to assessing code comprehension skills. Ultimately, these exercises show promise as test items, featuring mostly high discrimination as well as the ability to scale difficulty through the selection of the easier one-attempt or harder robustness (e.g. generate five times, not once) grading methods. Function naming exercises present high quality, easy to construct exam items. Further, they have exciting potential as a scaffolding step in the development of code comprehension, worthy of future study.

While future work should evaluate more code snippets for use as function naming exercises, to determine whether our findings generalize further, initial signs for this type of exercise are promising. Readers interested in deploying their own function naming exercises can use our feature in the \texttt{eiplgrader} package to ease adoption and deployment.

\bibliographystyle{ACM-Reference-Format}
\balance
\bibliography{sample-base}

\end{document}